\documentclass[aps, prc, 12pt, onecolumn, amsmath, amssymb, superscriptaddress, nofootinbib]{revtex4-2}
\usepackage{graphicx}
\usepackage{color}
\usepackage{booktabs}
\usepackage{lineno}

\newcommand{\be}{\begin{equation}}
\newcommand{\ee}{\end{equation}}
\newcommand{\beq}{\begin{eqnarray}}
\newcommand{\eeq}{\end{eqnarray}}
\newcommand{\ba}{\begin{array}}
\newcommand{\ea}{\end{array}}

\begin{document}

\title{Universal Mass Equation for \\ Equal-Quantum Excited-States Sets~III}

\vspace{3mm}
\date{\today}

\author{\mbox{L.~David~Roper}} 
\altaffiliation{Author: \texttt{roperld@vt.edu}} 
\affiliation{Prof. Emeritus of Physics, Virginia Polytechnic Institute and State University\\
    1001 Auburn Dr. SW, Blacksburg VA 24060, USA}

\author{\mbox{Igor~Strakovsky}} 
\altaffiliation{Author: \texttt{igor@gwu.edu}} 
\affiliation{Institute for Nuclear Studies, Department of Physics, 
    The George Washington University, Washington, DC 20052, USA} 

\begin{abstract}
Using our simple logarithmic Universal Mass Equation, we considered heavy non-$q\bar{q}$ meson states. PDG2026 reported a series of mesons with quantum numbers not allowed for standard $q\bar{q}$ states. Using LQCD predictions, we evaluated a scalar $0^{++}$ and tensor $2^{++}$ glueball series for which predictions are available. This approach allows us to predict heavy states that require future experimental studies. Additionally, we consider the possible misidentification of some meson states.
\end{abstract} 
\maketitle

\section{Introduction} \label{Sec:Intro}

 Sixty years ago, the invention of quantum chromodynamics (QCD), which explains the confinement of quarks in mesons and baryons and forms a large part of the Standard Model of elementary particles and forces, was fundamental to our understanding of particle physics~\cite{Gell-Mann:1964ewy, Zweig:1964jf}. QCD gives rise to the hadron spectrum, and many $q\bar{q}$ and $qqq$ states have been observed, as the recent Particle Data Group (PDG) reported~\cite{ParticleDataGroup:2026mpi}. 

 At first sight, the seemingly simple picture of hadrons emerging from QCD is, in fact, deceptive. The complex field-theoretical dynamics lead to a large variety of spectroscopic configurations. The basic constituents, quarks and gluons, interact in manifold ways, leading to a variety of intrinsic configurations that form the hadron's inner confined core. Illustrations, for instance, of elementary quark-gluon configurations of mesons: meson, tetraquark, and hadro-quarkonium; and baryons: baryon, hybrid-baryon, and pentaquark. The additional configuration is a glueball meson, \textit{i.e.}, an extremely exotic hadron consisting solely of self-interacting gluons. The search for such kinds of exotic states has a long history. 
 
In our previous publications~\cite{Roper:2024ovj, Roper:2025sli}(RS), we reported a simple logarithmic Universal Mass Equation (UME) that fits excited-state masses for increasing radial excitation and otherwise equal quantum indicators. The fits worked well for all baryon, meson, and some exotic categories where data were available two years ago~\cite{ParticleDataGroup:2024cfk}. Our Part~I~\cite{Roper:2024ovj} introduced a logarithmic formula, Eq.~(\ref{eq:eq0}), that described the spectra of baryons and mesons using data from the Particle Data Listings~\cite{ParticleDataGroup:2024cfk}. 
The UME is 
\begin{equation}
    M_n = \alpha~Ln(n) + \beta \>,
\label{eq:eq0}
\end{equation}
where $n$ is the radial quantum number, and $\alpha$ and $\beta$ are free parameters in a $\chi^2$ minimization procedure using Breit-Wigner (BW) PDG data~\cite{ParticleDataGroup:2024cfk}. We use a Naming Scheme for Hadrons following Ref.~\cite{Amsler:2024}. The parameter $\alpha$ is the logarithmic slope, and $\beta$ is the ground-state mass in the data set $\beta = M_1$, since $ \ ln (1) = 0$; however, in a $\chi^2$ fit, $\beta$ may vary slightly from $M_1$. Using PDG low-lying (including ground) states, one can predict high-lying states that PDG does not yet report -- ``missed'' states~\cite{Koniuk:1979vw}. 
Our Part~II~\cite{Roper:2025sli} used data sets with only two known excited states with masses $M_1$ and $M_2$ (duo sets), which have reasonably low measurement uncertainties. Here, the parameter $\beta$ is set to $M_1$ exactly.

Due to the high experimental activity by the ATLAS, BaBar, BES, Belle, CMS, COMPASS, LHCb, and SND collaborations, the PDG is non-stop collecting new resonances.  This is evident when comparing the two-year-old editions~\cite{ParticleDataGroup:2024cfk} and the fresh editions by 2026~\cite{ParticleDataGroup:2026mpi}. In this paper (Part~III), we will follow our recent studies~\cite{Roper:2024ovj, Roper:2025sli} and focus mainly on two types of exotics: tetraquarks and glueballs. 

\section{UME for New States}

Actually, the PDG2026 reported several bump-hunting results with mostly 
unknown quantum numbers. We will present results known by 
PDG2026~\cite{ParticleDataGroup:2026mpi} with quantum numbers: 
$I^G(J^{PC})$, where $I$ is isospin, $G$ is $G$-parity, $J$ is total 
angular momentum (spin), $P$ is parity, and $C$ is charge conjugation 
parity. We will focus on mesons whose properties are incompatible with a 
$q\bar{q}$ structure (exotic state)~\cite{Gutsch:2026}.

\subsection{UME for $T_{c\bar{c}\bar{s}1}$}

\begin{figure}[htb!]
\centering
{
        \includegraphics[width=0.7\textwidth,angle=0,keepaspectratio]{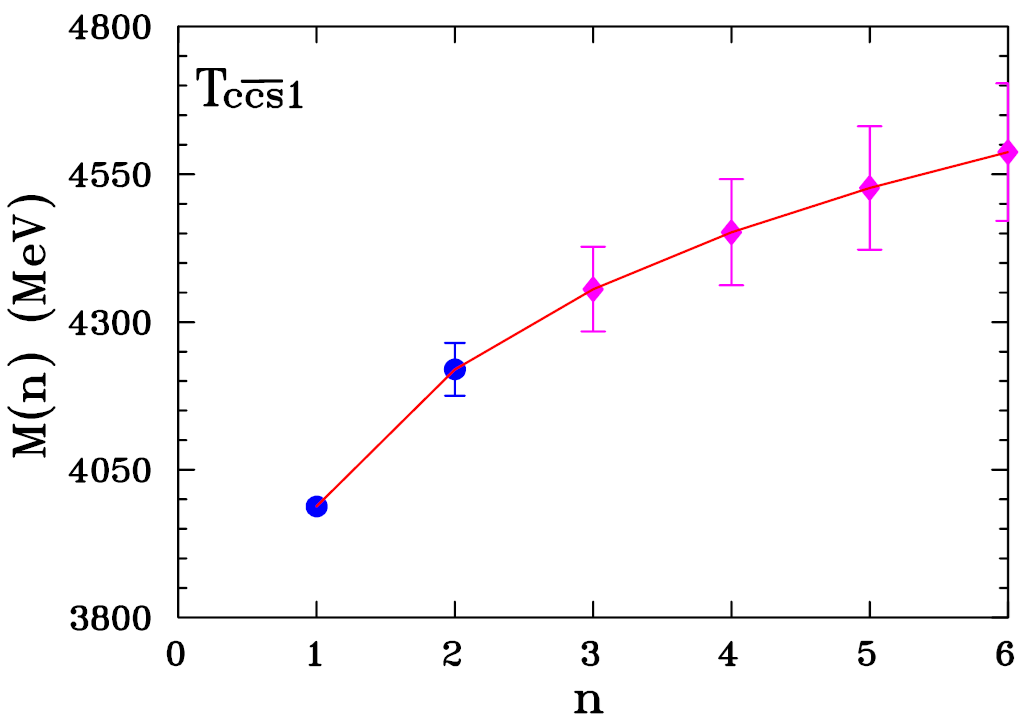}
}
\centerline{\parbox{1\textwidth}{
\caption[] {\protect\small
UME for $T_{c\bar{c}\bar{s}1}$. The first two blue filled circle data points are 
from~\cite{ParticleDataGroup:2026mpi}. The four magenta filled diamond points are projected excited states. The UME equation Eq.~(\ref{eq:eq0}) is $M_n= (335 \pm 65)~\mathrm{Ln(n)} + 3988~\mathrm{MeV}$. The red solid curve represents the best-fit result and its extrapolations for the four heavy-lying states.
}
\label{fig1}}}
\end{figure}

The properties of the $T_{c\bar{c}\bar{s}1}$ are incompatible with a 
$q\bar{q}$ structure (exotic state)~\cite{ParticleDataGroup:2026mpi}. UME results are shown in Fig.~\ref{fig1} and Table~\ref{tab1}. $I(J^P) = 1/2(1^+)$. Observed by the LHCb Collaboration in $B^+ \to T_{c\bar{c}\bar{s}1}(4000)^+ \phi$ with $T_{c\bar{c}\bar{s}1}
(4000)^+ \to J/\psi K^+$, using an amplitude analysis of $B^+ \to J/\psi \phi K^+$ with a significance of $15\sigma$~\cite{LHCb:2021uow}. The $J^P = 1^+$ assignment is favored with high significance. 
\begin{table}[htb!]
\centering \protect\caption{Best-fit results for $T_{c\bar{c}\bar{s}1}$. PDG2026 input: $T_{c\bar{c}\bar{s}1}(4000)$ and $T_{c\bar{c}\bar{s}1}(4220)$ came from Ref.~\cite{ParticleDataGroup:2026mpi}. The ground-state mass of $3988~\mathrm{MeV}$ could indicate that this is the $u\bar{s}c\bar{c}$ or $d\bar{s}c\bar{c}$ tetraquark with constituent mass $3922$ or $3926~\mathrm{MeV}$.
}

\vspace{2mm}
{%
\begin{tabular}{|c|c|c|}
\hline
PDG2026         &  Best-fit       & Comments  \tabularnewline
(GeV)           &  (GeV)          & \tabularnewline
\hline
3988.0$\pm$ 5.0 & 3988.0          & $Z_{cs}(4000)$ \tabularnewline
4220.0$\pm$45.0 & 4220.0$\pm$45.0 & $Z_{cs}(4220)^+$ \tabularnewline
                & 4355.7$\pm$71.3 & Predicted state \tabularnewline
                & 4452.0$\pm$90.0 & Predicted state \tabularnewline
                & 4526.7$\pm$104.5 & Predicted state \tabularnewline
                & 4587.7$\pm$116.3 & Predicted state \tabularnewline
\hline
\end{tabular}} \label{tab1}
\end{table}

\subsection{UME for $T_{c\bar{c}1}$}

\begin{figure}[htb!]
\centering
{
        \includegraphics[width=0.7\textwidth,angle=0,keepaspectratio]{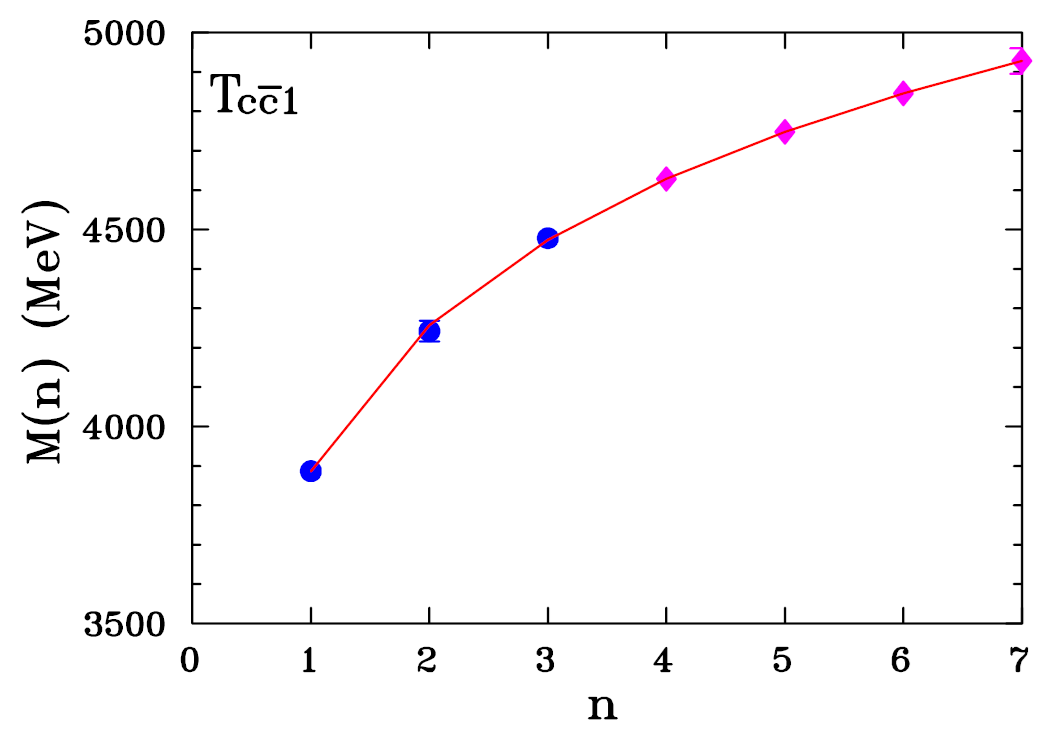}
}
\centerline{\parbox{1\textwidth}{
\caption[] {\protect\small
UME for $T_{c\bar{c}1}$. The first three blue-filled circle data points are 
from~\cite{ParticleDataGroup:2026mpi}. The four magenta-filled diamond points are projected excited states. The UME equation Eq.~(\ref{eq:eq0}) is $M_n= (520 \pm 14)~\mathrm{Ln(n)} + (3884 \pm 3)~\mathrm{MeV}$. The red solid curve shows the best fit and its extrapolations for the four heavy-lying states.
}
\label{fig2}}}
\end{figure}

The properties of the $T_{c\bar{c}1}$ are incompatible with a $q\bar{q}$ structure (exotic state)~\cite{ParticleDataGroup:2026mpi}. UME results are shown in Fig.~\ref{fig2} and Table~\ref{tab2}. $I^G(J^{PC}) = 1^+(1^{+ -})$. Charged low-lying $T_{c\bar{c}1}(3900)^\pm$ is seen as a peak in the invariant mass distribution of the $J/\psi \pi^\pm$ system by the BES~III Collaboration~\cite{BESIII:2013ris} in $e^+e^= \to \pi^+ \pi^- J/\psi$ with a CM energy of $4.26~\mathrm{GeV}$, and by radiative return from $e^+ e^-$ collisions at $W$ from 9.46 to 
$10.86~\mathrm{GeV}$ at Belle~\cite{Belle:2013yex}. Partial wave analysis of~\cite{BESIII:2017bua} determines $J^P = 1^+$ with more than $7\sigma$ significance. 
\begin{table}[htb!]
\centering \protect\caption{Best-fit results for $T_{c\bar{c}1}$. PDG2026 input: $T_{c\bar{c}1}$(3900), $T_{c\bar{c}1}(4200)$, and 
$T_{c\bar{c}1}(4430)$ came from Ref.~\cite{ParticleDataGroup:2026mpi}. The ground-state mass of $3887~\mathrm{MeV}$ could indicate that this is one of the three $u\bar{u}c\bar{c}\rightarrow d\bar{d}c\bar{c}$ tetraquarks with constituent masses $3772\rightarrow 3780~\mathrm{MeV}$. 
}

\vspace{2mm}
{%
\begin{tabular}{|c|c|c|}
\hline
PDG2026         &  Best-fit       & Comments  \tabularnewline
(GeV)           &  (GeV)          & \tabularnewline
\hline
3886.7$\pm$ 2.3 & 3886.7$\pm$ 2.3 & $Z_c(3900)$, $X(3900)$ \tabularnewline
4242.0$\pm$26.0 & 4257.4$\pm$12.9 & $Z_c(4200)$, $X(4200)^\pm$ \tabularnewline
4478.0$\pm$16.0 & 4474.3$\pm$19.1 & $Z_c(4430)$, $X(4430)^\pm$ \tabularnewline
                & 4628.2$\pm$23.5 & Predicted state \tabularnewline
                & 4747.6$\pm$26.9 & Predicted state \tabularnewline
                & 4845.1$\pm$29.7 & Predicted state \tabularnewline
                & 4927.5$\pm$32.1 & Predicted state \tabularnewline
\hline
\end{tabular}} \label{tab2}
\end{table}

\subsection{UME for Tetraquarks (Example: $T_{cc\bar{c}\bar{c}}$)}

\begin{figure}[htb!]
\centering
{
        \includegraphics[width=0.7\textwidth,angle=0,keepaspectratio]{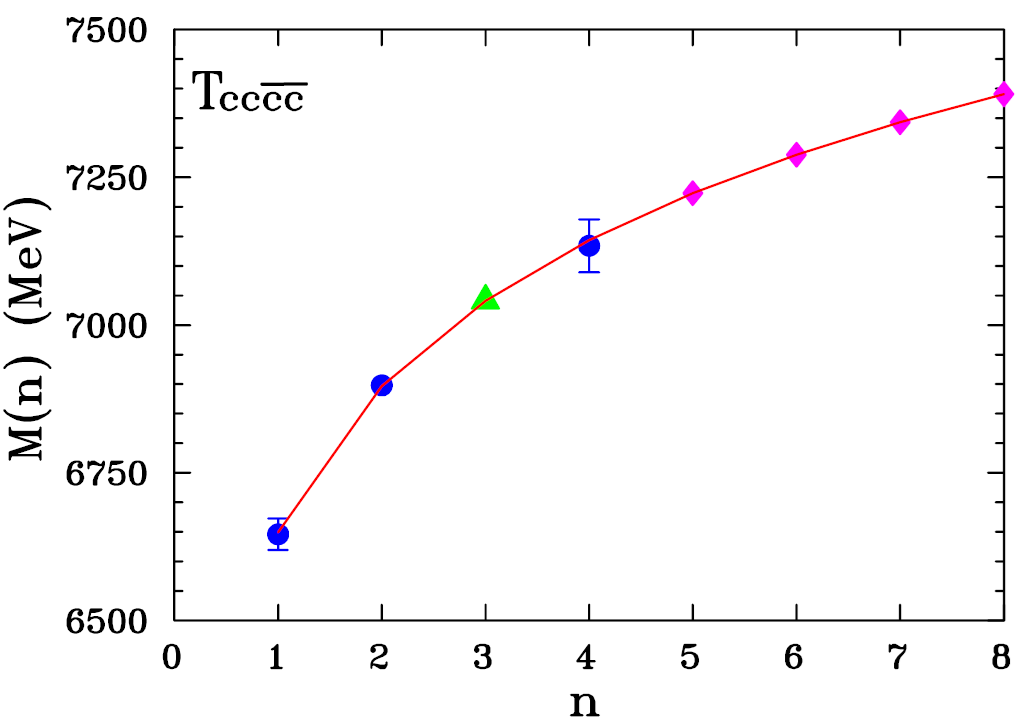}
}
\centerline{\parbox{1\textwidth}{
\caption[] {\protect\small
UME for $T_{cc\bar{c}\bar{c}}$ tetraquark. The first two and fourth blue-filled circle data points are 
from~\cite{ParticleDataGroup:2026mpi}. The green-filled triangle point is a missing state, and the four magenta-filled diamond points are projected excited states. The UME equation Eq.~(\ref{eq:eq0}) is $M_n= (357 \pm 34)~\mathrm{Ln(n)} + (6649 \pm 24)~\mathrm{MeV}$. The red solid curve shows the best fit and its extrapolations for the four heavy-lying states.
}
\label{fig3}}}
\end{figure}

It is not surprising that UME works well for tetraquarks, as they are essentially double mesons. An example tetraquark, $T_{cc\bar{c}\bar{c}}$, is shown in Fig.~\ref{fig3} and 
Table~\ref{tab3}. $I^G(J^{PC}) = 0^+(2^{+ +})$. CMS Collaboration determined those to $J^{PC} = 2^{+ +}$~\cite{CMS:2025fpt}. Obviously, 
these states are incompatible with a $q\bar{q}$ structure~\cite{ParticleDataGroup:2026mpi}. 
\begin{table}[htb!]
\centering \protect\caption{Best-fit results for $T_{cc\bar{c}\bar{c}}$. PDG2026 input: $T_{cc\bar{c}\bar{c}}(6600)$, $T_{cc\bar{c}\bar{c}}(6900)$, and $T_{cc\bar{c}\bar{c}}(7100)$ came from Ref.~\cite{ParticleDataGroup:2026mpi}.
}

\vspace{2mm}
{%
\begin{tabular}{|c|c|c|}
\hline
PDG2026         &  Best-fit       & Comments  \tabularnewline
(GeV)           &  (GeV)          & \tabularnewline
\hline
6646.0$\pm$26.5 & 6649.4$\pm$16.4 & \tabularnewline
6898.0$\pm$12.0 & 6896.6$\pm$16.2 & $X(6900)$ \tabularnewline
                & 7041.2$\pm$16.1 & Missed state \tabularnewline
7134.0$\pm$44.5 & 7143.7$\pm$15.9 & \tabularnewline
                & 7223.3$\pm$15.9 & Predicted state \tabularnewline
                & 7288.3$\pm$15.8 & Predicted state \tabularnewline
                & 7343.3$\pm$15.8 & Predicted state \tabularnewline
                & 7390.9$\pm$15.7 & Predicted state \tabularnewline
\hline
\end{tabular}} \label{tab3}
\end{table}

There are several methods to calculate tetraquark masses (Table~\ref{tab:TQ_methods}). Some of them are in reasonable agreement with our picture.
\begin{table*}[htb!]
\centering \protect\caption{Different ways to calculate tetraquark masses.
}

\vspace{2mm}
    \centering
    \begin{tabular}{|c|c|}\hline
    \textbf{Method }  &\textbf{Basic Principle} \tabularnewline
\hline
Lattice QCD                 & Calculate quark and gluon interactions \tabularnewline
& on a spacetime lattice~\cite{Luscher:1985dn} \tabularnewline
Constituent-quark models    & Effective quark masses and interactions~\cite{Jaffe:1976ig} \tabularnewline
Diquark–anti-diquark models & Tetraquarks as two interacting clusters~\cite{Maiani:2004vq} \tabularnewline
Hadronic molecular models   & Tetraquarks as bound or resonant pairs of mesons~\cite{Tornqvist:1993ng} \tabularnewline
QCD sum rules               & Relate QCD correlation functions to hadron masses~\cite{Shifman:1978bx} \tabularnewline
Bethe–Salpeter equations    & Relativistic bound-state equations~\cite{Eichmann:2015cra} \tabularnewline
Potential and Schr\"odinger models & Effective four-body or two-cluster equations~\cite{Vijande:2006jf} \tabularnewline
Born–Oppenheimer methods    & Separate heavy-quark and light-field dynamics~\cite{Braaten:2014qka} \tabularnewline
\hline
    \end{tabular}
    \label{tab:TQ_methods}
\end{table*}

\subsection{UME for $T_{b\bar{b}1}$}

\begin{figure}[htb!]
\centering
{
        \includegraphics[width=0.7\textwidth,angle=0,keepaspectratio]{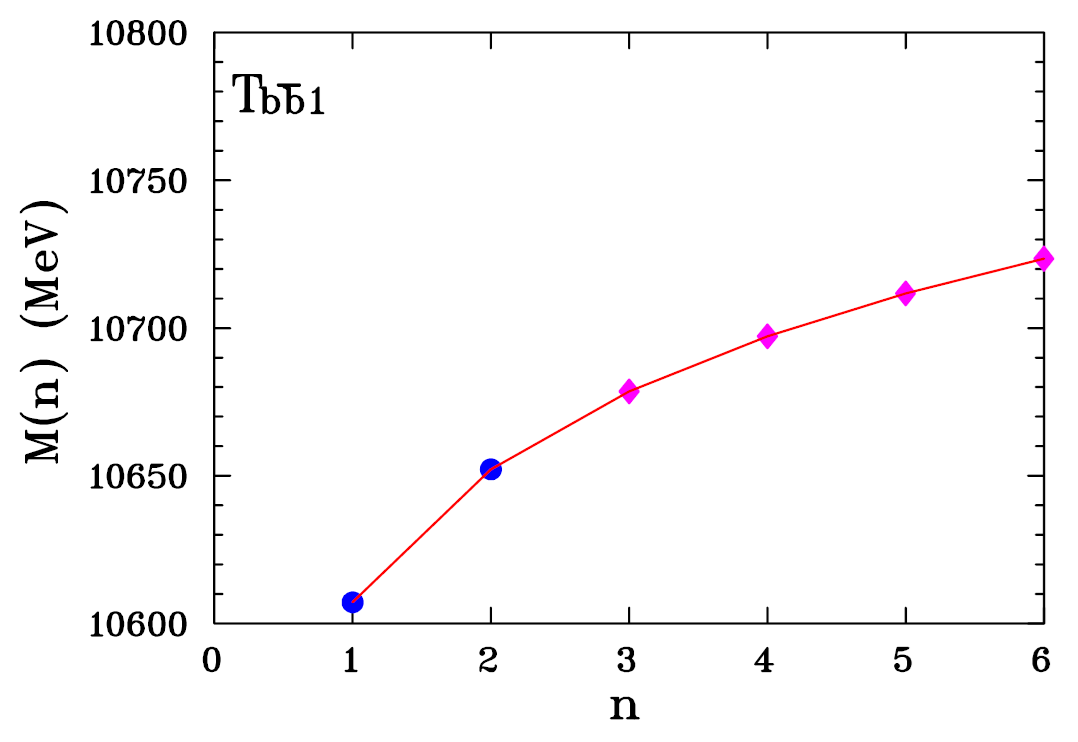}
}
\centerline{\parbox{1\textwidth}{
\caption[] {\protect\small
UME for $T_{b\bar{b}1}$. The first two blue-filled circle data points are 
from~\cite{ParticleDataGroup:2026mpi}. The four magenta-filled diamond points are projected excited states. The UME equation Eq.~(\ref{eq:eq0}) is $M_n= (65 \pm 2)~\mathrm{Ln(n)} + 10607.2~\mathrm{MeV}$. The red solid curve shows the best fit and its extrapolations for the four heavy-lying states.
}
\label{fig4}}}
\end{figure}

The properties of the $T_{b\bar{b}1}$ are incompatible with a $q\bar{q}$ structure (exotic state)~\cite{ParticleDataGroup:2026mpi}. UME results are shown in Fig.~\ref{fig4} and Table~\ref{tab4}. $I^G(J^{PC}) = 1^+(1^{+ -})$. Observed by the Belle Collaboration~\cite{Belle:2011aa} in 
$\Upsilon(5S)$ decays to $\Upsilon(nS) \pi^+ \pi^-$, where (n =
1, 2, 3) and $h_b(mP) \pi^+ \pi^-$ (m = 1, 2). $J^P = 1^+$ is favored based on angular analysis.
\begin{table}[htb!]
\centering \protect\caption{Best-fit results for $T_{b\bar{b}1}$. PDG2026 input: $T_{b\bar{b}1}(10610)$ and $T_{b\bar{b}1}(10650)$ came from Ref.~\cite{ParticleDataGroup:2026mpi}. The ground-state mass of $10607~\mathrm{MeV}$ could indicate that this is the $s\bar{s}b\bar{b}$  tetraquark with constituent mass $10432~\mathrm{MeV}$.
}

\vspace{2mm}
{%
\begin{tabular}{|c|c|c|}
\hline
PDG2026         &  Best-fit       & Comments  \tabularnewline
(GeV)           &  (GeV)          & \tabularnewline
\hline
10607.2$\pm$2.0 & 10607.2         & $Z_b(10610)$, $X(10610)$ \tabularnewline
10652.2$\pm$1.5 & 10652.2$\pm$1.5 & $Z_b(10650)$, $X(10650)^\pm$ \tabularnewline
                & 10678.5$\pm$2.4 & Predicted state \tabularnewline
                & 10697.2$\pm$3.0 & Predicted state \tabularnewline
                & 10711.7$\pm$3.5 & Predicted state \tabularnewline
                & 10723.5$\pm$3.9 & Predicted state \tabularnewline
\hline
\end{tabular}} \label{tab4}
\end{table}

\subsection{UME for Glueballs}

\begin{figure}[htb!]
\centering
{
        \includegraphics[width=0.7\textwidth,angle=0,keepaspectratio]{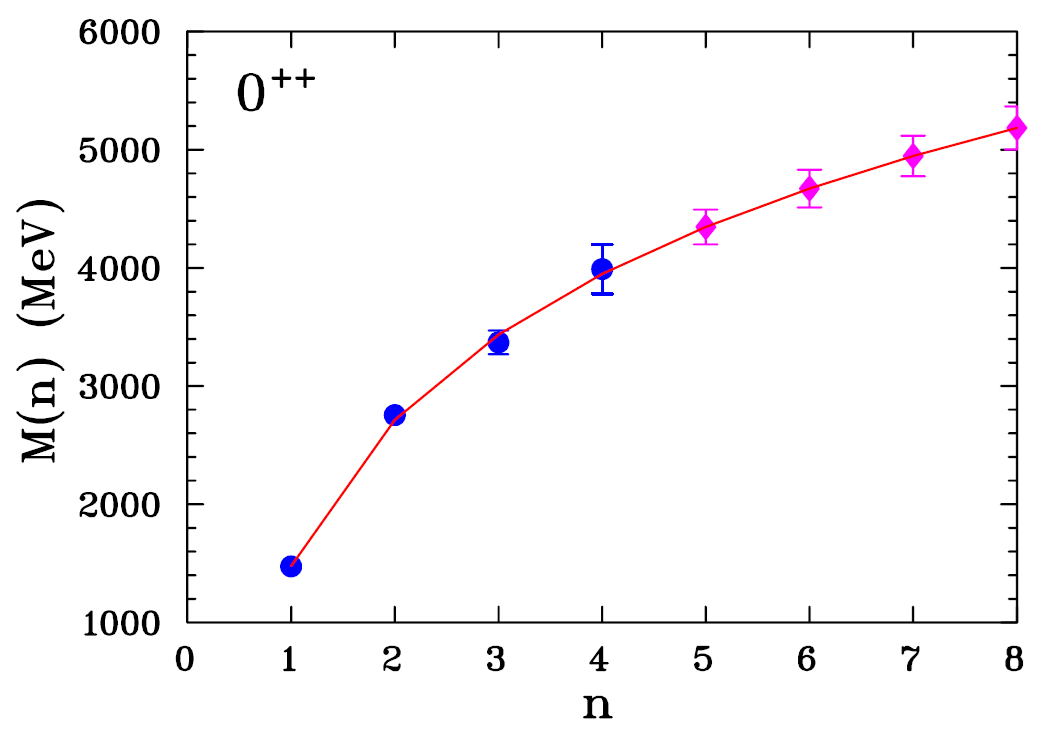}
}
\centerline{\parbox{1\textwidth}{
\caption[] {\protect\small
UME for scalar $0^{++}$ glueballs. The first four blue-filled circle data points are from LQCD calculations~\cite{Meyer:2004gx}. Four magenta-filled diamond projected masses are shown. The UME equation Eq.~(\ref{eq:eq0})is $M_n= (1709 \pm 70)~\mathrm{Ln(n)} + (1449 \pm 30)~\mathrm{MeV}$. The red solid curve shows the best fit and its extrapolations for the four heavy-lying cases.
 }
\label{fig5}}}
\end{figure}

\begin{figure}[htb!]
\centering
{
        \includegraphics[width=0.7\textwidth,angle=0,keepaspectratio]{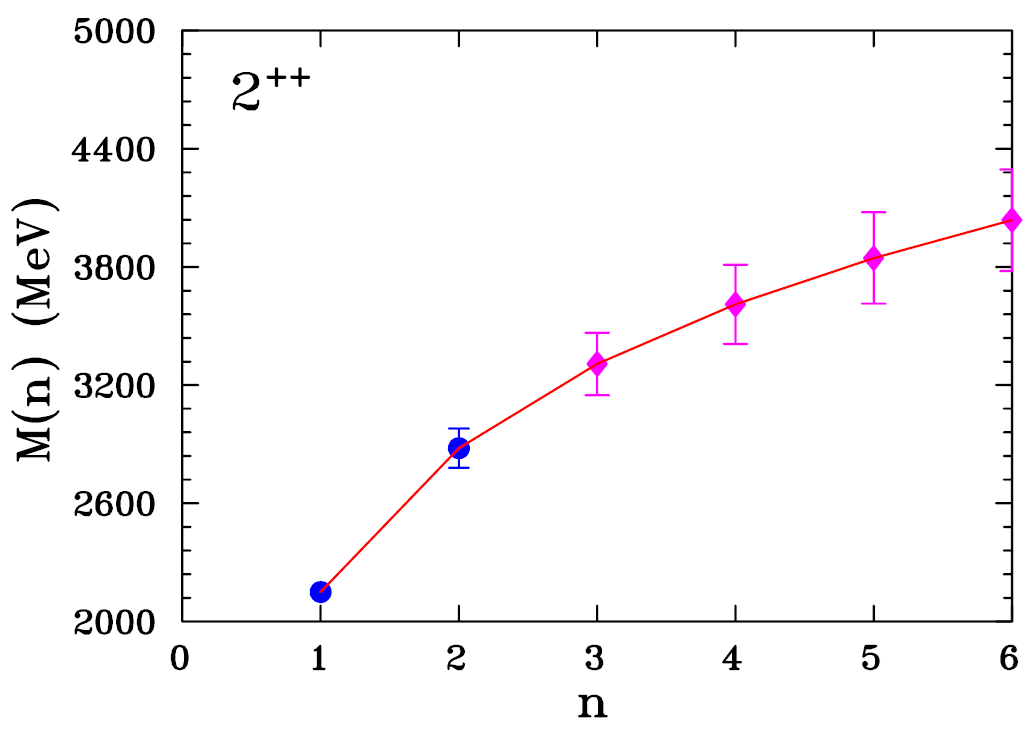}
}
\centerline{\parbox{1\textwidth}{
\caption[] {\protect\small
UME for tensor $2^{++}$ glueballs. The first four blue-filled circle data points are from LQCD calculations~\cite{Meyer:2004gx}. Four magenta-filled diamond projected masses are shown. The UME equation Eq.~(\ref{eq:eq0})is $M_n= (1053 \pm 144)~\mathrm{Ln(n)} + 2150~\mathrm{MeV}$. The red solid curve shows the best fit and its extrapolations for the four heavy-lying cases.
 }
\label{fig6}}}
\end{figure}

Glueballs are difficult to identify experimentally because they have masses similar to those of the constituent-quark states.  As discussed below, perhaps some glueball states have been misidentified as quark states. The lowest mass glueballs are digluons ($gg$) and trigluons ($ggg$) (Table~\ref{tab:gg/ggg}). There are several calculation methods to determine glueball masses (Table~\ref{tab:GB_methods}). For our study, we selected two series of glueballs predicted by LQCD~\cite{Meyer:2004gx}: four scalar $0^{++}$s and two tensor $2^{++}$s. Mixing effects are already essential in the pure gauge case and lead to a large intercept of a pomeron-like trajectory. UME results for scalar $0^{++}$ are shown in Fig.~\ref{fig5} and Table~\ref{tab5} and for tensor $2^{++}$ are shown in Fig.~\ref{fig6} and Table~\ref{tab6}. The potential candidates for scalar glueballs are $f_0(1500)$ and $f_0(1710)$, while the potential candidates for tensor glueballs are $f_J(2220)$, $f_2(2300)$, and $f_2(2340)$. 
\begin{table}[htb!]
\centering \protect\caption{Features of $gg$ and $ggg$ glueballs.
}

\vspace{2mm}
    \centering
    \begin{tabular}{|c|c|c|} 
\hline
Property                     &  $gg$& $ggg$ \tabularnewline
\hline
Gluon color states available &  8   & 8 \tabularnewline
Independent color singlets   &  1   & 2 \tabularnewline
Charge conjugation           & C=+1 & C= +1, -1 \tabularnewline
$J^{PC}$                     &  Yes & Yes \tabularnewline
Radial excited states        &  Yes & Yes \tabularnewline
\hline
    \end{tabular}
    \label{tab:gg/ggg}
\end{table}
\begin{table*}[htb!]
\centering \protect\caption{There are several different ways to calculate glueball masses.
}
\label{tab:GB_methods}

\vspace{2mm}
\begin{tabular}{|c|c|}
\hline
\textbf{Method}          & \textbf{Basic principle} \tabularnewline
\hline
Lattice QCD              & QCD on a spacetime lattice~\cite{Michael:1996ek} \tabularnewline
QCD sum rules            & Relate gluonic correlation functions to
hadron properties~\cite{Shifman:1978bx} \tabularnewline
Constituent-gluon models & Glueballs as bound states
of effective massive gluons~\cite{Cornwall:1981zr} \tabularnewline
Potential models         & Bound-state equations using
effective gluon interactions~\cite{Cornwall:1982zn} \tabularnewline
Bethe–Salpeter equations & Relativistic bound-state equations for
interacting gluons~\cite{Heupel:2012ua} \tabularnewline
Flux-tube/string models  & Gluonic fields as closed loops or
strings~\cite{Isgur:1984bm} \tabularnewline
Holographic QCD          & Higher-dimensional gravitational dual \tabularnewline
 & Broad patterns of glueball spectra~\cite{Witten:1998zw} \tabularnewline
Functional QCD methods   & Truncated Dyson-Schwinger and
related equations~\cite{Fischer:2006vf} \tabularnewline
\hline
\end{tabular}
\end{table*}

Meyer, in his 2005 Dr. Phil. Thesis at the University of Oxford~\cite{Meyer:2004gx}, used lattice QCD to calculate several series of glueballs.
This UME fit for an equal-quantum set of four glueballs ($0^{++}$ sample) shows that the UME works for glueballs.

\begin{table}[htb!]
\centering \protect\caption{Best-fit results for scalar $0^{++}$. LQCD input came from Ref.~\cite{Meyer:2004gx}.
}

\vspace{2mm}
{%
\begin{tabular}{|c|c|c|}
\hline
LQCD         &  Best-fit    & Comments  \tabularnewline
(GeV)        &  (GeV)       & \tabularnewline
\hline
1475$\pm$30  & 1478$\pm$29  & \tabularnewline
2755$\pm$70  & 2713$\pm$80  & \tabularnewline
3370$\pm$100 & 3436$\pm$110 & \tabularnewline
3990$\pm$210 & 3949$\pm$131 & \tabularnewline
             & 4346$\pm$147 & Predicted state \tabularnewline
             & 4671$\pm$161 & Predicted state \tabularnewline
             & 4946$\pm$172 & Predicted state \tabularnewline
             & 5184$\pm$182 & Predicted state \tabularnewline
\hline
\end{tabular}} \label{tab5}
\end{table}
\begin{table}[htb!]
\centering \protect\caption{Best-fit results for tensor $2^{++}$. LQCD input came from Ref.~\cite{Meyer:2004gx}.
}

\vspace{2mm}
{%
\begin{tabular}{|c|c|c|}
\hline
LQCD         &  Best-fit    & Comments  \tabularnewline
(GeV)        &  (GeV)       & \tabularnewline
\hline
2150$\pm$30  & 2150         & \tabularnewline
2880$\pm$100 & 2880$\pm$100 & \tabularnewline
             & 3307$\pm$159 & Predicted state \tabularnewline
             & 3610$\pm$200 & Predicted state \tabularnewline
             & 3845$\pm$232 & Predicted state \tabularnewline
             & 4037$\pm$259 & Predicted state \tabularnewline
\hline
\end{tabular}} \label{tab6}
\end{table}

This UME is fit for an equal-quantum set of four scalar glueballs ($0^{++}$ and two tensor glueballs ($2^{++}$), showing that the UME works for glueballs.
Note that the glueball UME $\alpha$ parameter (the logarithmic slope in Eq.~(\ref{eq:eq0})) is about 3.5 times larger than $\alpha$ for tetraquarks.

\section{Possible Misidentification of Glueball or Tetraquark States}

RS states the following: ``Several states reported by PDG2026 are out of the sequence $f_2(1430)$, $f_2(1571)$, $f_2(2157)$, $f_2(2297)$, and $f_2(2346)$'' and ``$f_0$ has two states reported in the PDG2026  with masses $1784\pm 15~\mathrm{MeV}$ and $2470\pm 7~\mathrm{MeV}$, which do not fit 
Eq.~(\ref{eq:eq0}).'' ``Several states reported by PDG2026 are out of the UME sequence $\eta(1294)$, $\eta(1409)$, and $\eta(2221)$.''

This section matches, within uncertainties, the known glueballs discussed above; three such matches exist. Then, tetraquarks are matched with reasonably negative constituent-quark binding energies; four such matches exist. See Table~\ref{misID}

\begin{table*}[htb!]
\centering
\caption{Tetraquark and glueball near matches to possibly misidentified mesons.``Misnomer'' is a ``misapplied or inappropriate name,'' TQ is tetraquark, $M_c$ is constituent quark mass, GB is glueball ($gg$ and $ggg$), $M_{GB}$ is glueball mass, HM is Ref.~\cite{Meyer:2004gx}, and AT 
is Ref.~\cite{Athenodorou:2020ani}.  
}
\label{misID}
\vspace{2mm}
\begin{tabular}{|c|c|c|c|c|c|}
\hline
Misnomer~? & $\eta$ & $\eta$ & $f_2$ & $f_2$ & $f_0$ \\ \hline
$I^G(J^{PC})$ & $0^+(0^{- +})$ & $0^+(0^{- +})$ & $0^+(2^{++})$ & $0^+(2^{++})$ & $0^+(0^{++})$ \\ \hline
Mass (MeV) & 1294$\pm$4 & 1408$\pm$1.5 & 1440$\pm$11 & 1571$\pm$13 & 1785$\pm$13 \\ \hline
TQ &   & T($u\bar{u}u\bar{u}$) & T($u\bar{u}d\bar{d}$) & T($d\bar{d}d\bar{d}$) & T($u\bar{u}s\bar{s}$) \\ \hline
$M_c$ (MeV)   &  & 1344 & 1352 & 1360 & 1644 \\ \hline
$M_c-M$ (MeV) &  & -64 & -88 & -211 & -141  \\ \hline \hline
Misnomer~? & $f_2$ & $\eta$ & $f_2$ & $f_2$ & $f_0$ \\ \hline
$I^G(J^{PC})$ & $0^+(2^{++})$ & $0^+(0^{- +})$ 
& $0^+(2^{++})$ & $0^+(2^{++})$ & $0^+(0^{++})$ \\ \hline
Mass (MeV)& 2157$\pm$12& 2221$\pm$12& 2297$\pm$28 & 2346$\pm$15 & 2470$\pm$6 \\ \hline
GB & HM GB $2^{++}$ & HM GB $0^{–+}$ &   & AT $gg$ $2^{++}$ & \\ \hline
$M_{GB}$ (MeV)   & 2150$\pm$30 & 2250$\pm$60 & & 2376$\pm$32 & \\ \hline
\end{tabular}
\end{table*}

\section{Conclusion}

Following our previous RS papers, we considered the case of exotic mesons, heavy non-$q\bar{q}$ mesons. Obviously, our simple logarithmic Universal Mass Equation works and can predict heavy-lying states. We hope that experimentalists will have a chance to use this guide to look for new states that the PDG missed.

\section{Acknowledgments}

This work was supported in part by the U.S. Department of Energy, Office of Science, Office of Nuclear Physics, under Award Number DE--SC0016583. 


\end{document}